\documentclass[conference,compsoc]{IEEEtran}
\usepackage{url,hyperref,microtype}
\usepackage{cases}
\usepackage{graphicx}

\ifCLASSOPTIONcompsoc
  \usepackage[nocompress]{cite}
\else
  \usepackage{cite}
\fi
\ifCLASSINFOpdf
\else
\fi
\begin{document}
%
\title{TrueHappiness: \\
Neuromorphic Emotion Recognition on TrueNorth}



%
\author{\IEEEauthorblockN{Peter U. Diehl\IEEEauthorrefmark{1}$^{1}$,
Bruno U. Pedroni\IEEEauthorrefmark{2}$^{1}$,
Andrew Cassidy\IEEEauthorrefmark{3}, 
Paul Merolla\IEEEauthorrefmark{3},
Emre Neftci\IEEEauthorrefmark{2}\IEEEauthorrefmark{4} and
Guido Zarrella\IEEEauthorrefmark{5}}
\IEEEauthorblockA{\IEEEauthorrefmark{1}Institute of Neuroinformatics\\
ETH Zurich and University Zurich, Switzerland\\ 
Email: peter.u.diehl@gmail.com}
\IEEEauthorblockA{\IEEEauthorrefmark{2}Institute for Neural Computation, UC San Diego, La Jolla, USA\\
Email: bpedroni@eng.ucsd.edu}
\IEEEauthorblockA{\IEEEauthorrefmark{3}IBM Research Almaden, San Jose, CA, USA\\}
\IEEEauthorblockA{\IEEEauthorrefmark{4}Department of Cognitive Sciences, UC Irvine, Irvine, USA}
\IEEEauthorblockA{\IEEEauthorrefmark{5}The MITRE Corporation, Bedford, MA, USA}
$^1$ Peter U. Diehl and Bruno U. Pedroni have contributed equally to this work}


\maketitle

\begin{abstract}
We present an approach to constructing a neuromorphic device that responds to language input by producing neuron spikes in proportion to the strength of the appropriate positive or negative emotional response. 
Specifically, we perform a fine-grained sentiment analysis task with implementations on two different systems: one using conventional spiking neural network (SNN) simulators and the other one using IBM's Neurosynaptic System TrueNorth. 
Input words are projected into a high-dimensional semantic space and processed through a fully-connected neural network (FCNN) containing rectified linear units trained via backpropagation.
After training, this FCNN is converted to a SNN by substituting the ReLUs with integrate-and-fire neurons.
We show that there is practically no performance loss due to conversion to a spiking network on a sentiment analysis test set, i.e. correlations between predictions and human annotations differ by less than 0.02 comparing the original DNN and its spiking equivalent.
Additionally, we show that the SNN generated with this technique can be mapped to existing neuromorphic hardware -- in our case, the TrueNorth chip.
Mapping to the chip involves 4-bit synaptic weight discretization and adjustment of the neuron thresholds.
The resulting end-to-end system can take a user input, i.e. a word in a vocabulary of over 300,000 words, and estimate its sentiment on TrueNorth with a power consumption of approximately 50 $\mu W$.
\end{abstract}


%
\IEEEpeerreviewmaketitle

\section{Introduction}
The rise of the internet and more powerful computers has enabled an unprecedented ability to interpret and understand natural language.
Deep neural networks (DNN) can be trained on massive datasets to perform a wide variety of natural language understanding and generation tasks \cite{collobert2011, mikolov2013, sutskever2014}.
One drawback of DNNs is that they usually require power hungry hardware, such as GPUs, posing a problem for mobile devices (e.g. smartphones), which present very stringent power constraints.
A common solution is to outsource the computation to the cloud by sending the data to a data center, processing it there and then sending the results back to the mobile device.
This works well as long as the amount of data to be processed is limited and as long as there is a reliable connection between the data center and the mobile device.
If either one of these conditions is not met, the system will not operate adequately and the user is left without the desired functionality.

A possible solution for the problem of high power consumption is to use neuromorphic hardware to perform the processing \cite{Indiveri_etal06, Khan_etal08, Schemmel_etal10, Indiveri_etal11}.
These brain-inspired systems work on an extremely low power budget: for example, the IBM's Neurosynaptic System TrueNorth can simulate 1 million neurons using less than 100 $mW$ \cite{Merolla_etal14}.
Mapping DNNs to neuromorphic hardware would make viable pattern recognition systems which present simultaneously low power and high performance \cite{Esser_etal13, marti2015}.
As a practical example, a TrueNorth chip analyzing a stream of language content could run on an iPhone battery nonstop for a week. 

Here we present a first example of a NLP system that is based on spiking neural networks and is also implemented on neuromorphic hardware.
Specifically, the system performs fine-grained sentiment analysis, i.e. evaluating how positive/negative a word or phrase is on a 1 to 9 scale.
We start by training a DNN with rectified linear units (ReLU) on a dataset labeled with crowd-sourced happiness ratings \cite{dodds2011}.
After training, we substitute the ReLUs with integrate-and-fire neurons and adjust neuron thresholds and scale the weights appropriately \cite{Diehl_etal15}. 
A comparison between the original DNN and the converted SNN on the sentiment analysis test set shows that the drop in correlation  between prediction and target due to the conversion to a spiking network is less than 0.02 for any of the tested setups.
Using the SNN, we provide code for a real-time interactive demo, where the user can query any word in a $>$300,000 word vocabulary and compute the associated sentiment estimate.

In addition to the SNN that is simulated on a traditional computer, we present an implementation of the same fine-grained sentiment analysis task on TrueNorth.
For the construction of this network we start from the generic SNN and proceed by mapping the synaptic weights between neurons using a quantization strategy with resulting weight precision of 4 bits. 

We call this approach ``train-then-constrain'', comparing it to the ``constrain-then-train'' used in \cite{Esser_etal2015}. 
Using the ``linear reset'' mode, TrueNorth neurons can be configured to produce spiking rates similar to ReLUs.
The final network uses 3 cores, consuming less than 0.1\% of a TrueNorth chip to process language with an estimated power consumption of less than 50 $\mu W$.

The result is an important first step in the creation of a new generation of spiking cognition systems: a neuromorphic device which receives words as input and outputs a variable number of spikes in proportion to the strength of expected emotional response.

\section{Methods}
\begin{figure}
\begin{center}
\includegraphics[width=8.5cm]{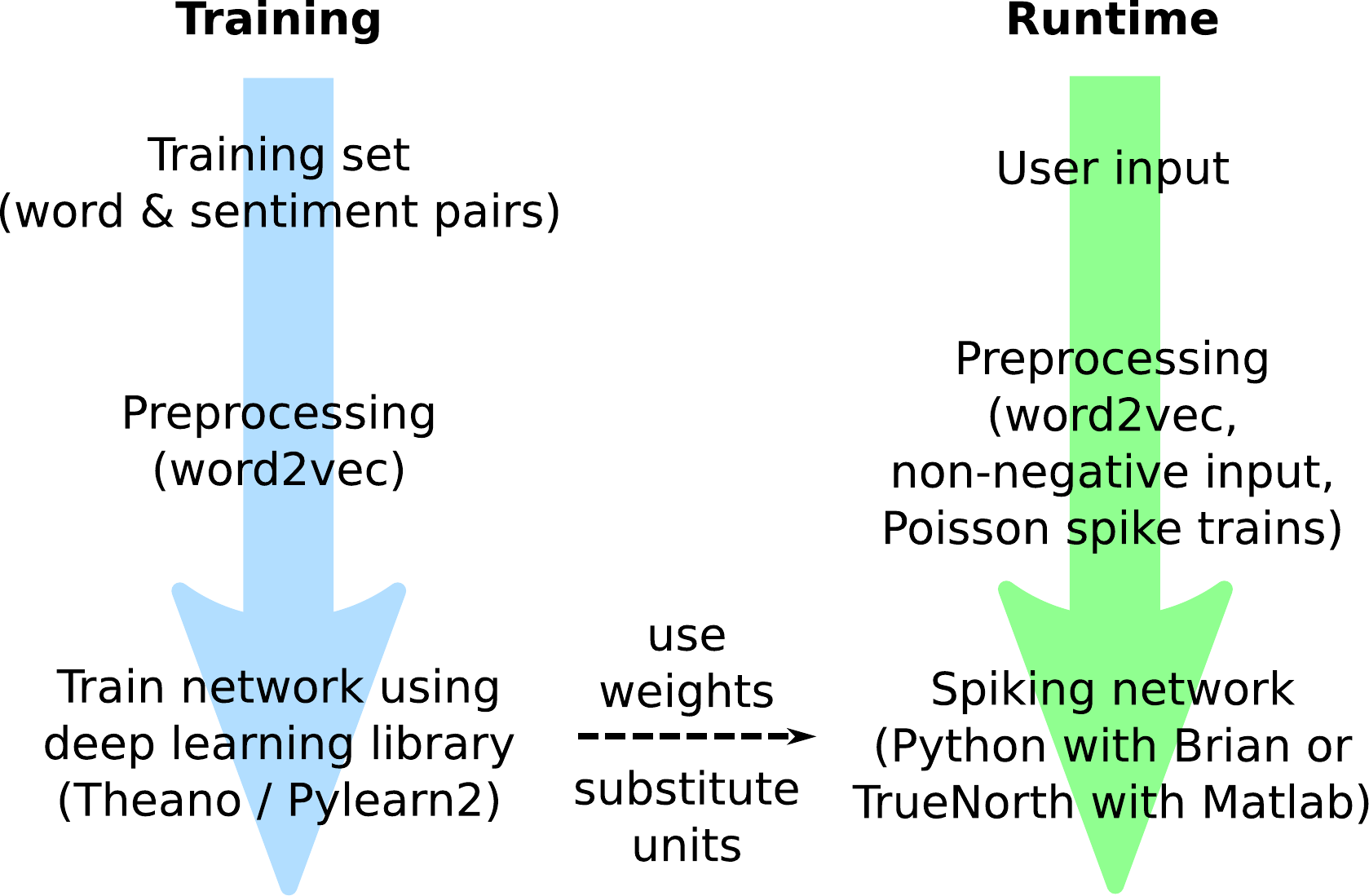}
\end{center}
\label{fig:conversion_process}
\caption{Conversion of the fully-connected neural network to a spiking neural network and a TrueNorth compatible network. The necessary steps are 1) training of the network using a traditional deep-learning library 2) substitution of the ReLU units with integrate-and-fire neurons 3) making inputs non-negative by either doubling number of input units or applying the exponential function 4) converting the real-numbered inputs to Poisson spike-trains. Additionally, for the TrueNorth network the weights are discretized to 4-bit precision and unit thresholds are adapted.}
\end{figure}

The workflow is depicted in figure \ref{fig:conversion_process}.
We begin by learning distributed word representations using the word2vec method, which models word meaning via a training process that predicts word co-occurrences in an unlabeled text corpus.
We then convert the words in our training set into vectors, which we use as input features in the next processing stage where a neural network learns to predict the sentiment score associated with the input word.
Training is done using traditional backpropagation.

After training is completed, the learned weights are used to construct a spiking neural network that performs the same task\footnote{The code for running the python based demo of TrueHappiness is available under https://github.com/peter-u-diehl/truehappiness.}.

Finally, this spiking network is mapped onto TrueNorth.
This is a "train-then-constrain" methodology, where an unconstrained network is learned using full precision weights, values, and neuron types, and then converted to a hardware compatible network.  
This contrasts with a "constrain-then-train" approach that accounts for the architectural constraints directly in the learning algorithm \cite{Esser_etal2015}.

\subsection{Data and Pretraining}
The dataset used here was constructed from data first presented in \cite{dodds2011}.
It contains 4500 words with corresponding crowdsourced sentiment labels, i.e. a number between 1 and 9 (inclusive).

Our goal is to learn a fine-grained sentiment predictor that generalizes to words and phrases that are not in the training set. 
As such we need a large dictionary of words encoded with a feature representation that captures latent attributes of word meaning and usage.
For this purpose we pre-train a word2vec \cite{mikolov2013} model which projects words into a high-dimensional space where the location of the words corresponds to their meaning.
Word similarities can be computed using the cosine distance between word vectors, e.g. "good" is close to "great" in this space but far away from "terrorist".
It is also possible to use vector arithmetic to compute analogies and create features for novel or ambiguous concepts.
For example, in our model the vector calculated from wordvector('happy') - wordvector('laughter') + wordvector('sad') is nearby wordvector('anguish').
A downstream model trained to recognize the sentiment of 'happy', 'laughter', and 'sad' could thus infer without explicit training that 'anguish' is a strongly negative term.

Our word vector representations are learned from a large corpus, in particular the text of the English Wikipedia. 
After some preprocessing of the 3.4 billion tokens in the corpus, like substitution of uncommon words with an "unknown" token, we use word2vec's skipgram algorithm with negative sampling\cite{mikolov2013} to learn a vector for the most common 324,263 words.
Each word is thus represented as a vector of 64 dimensions, a number selected for convenience in mapping to the TrueNorth architecture (explained in section \ref{sec:implementation_details}).

These word vectors form the basis for the inputs to a fully-connected feedforward neural network which we train to predict fine-grained sentiment scores.
The network contains a hidden layer of 64 rectified linear units without biases and a single output unit connected by linear weights. 
The weights were trained via stochastic gradient descent using 4000 labeled examples.
Hyperparameters were tuned to optimize performance on a held-out evaluation set of 500 examples. 

\begin{figure*}[t]
\begin{center}
\includegraphics[width=17.5cm]{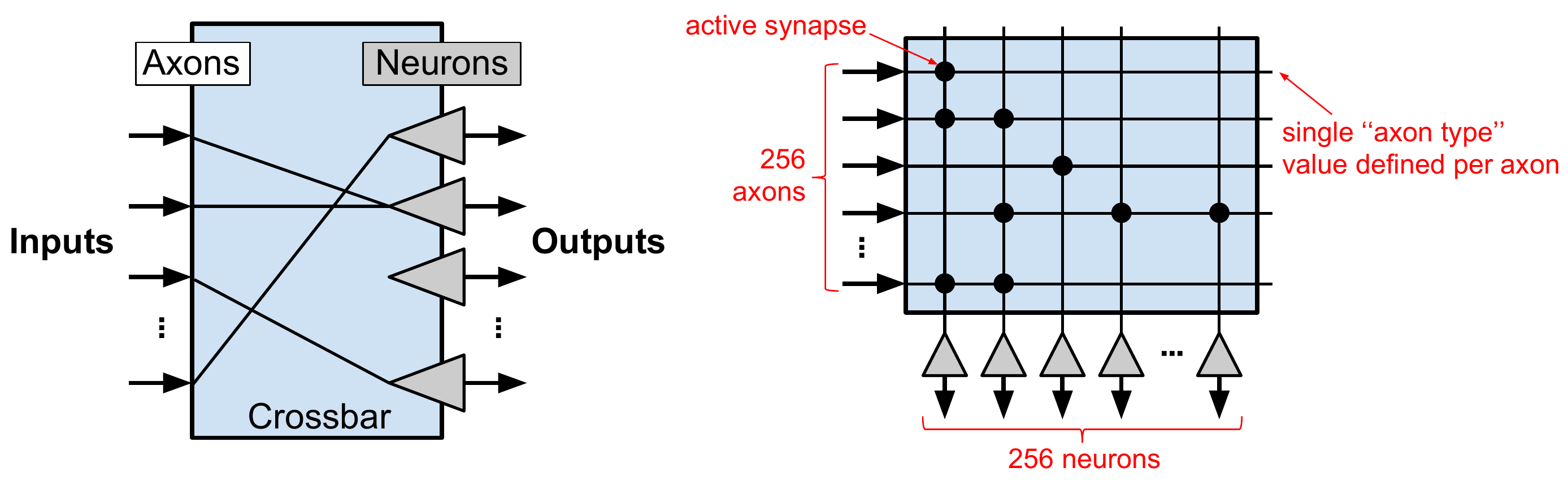}
\caption{High-level abstraction of a TrueNorth core: the axons can be seen as the inputs, while the neurons integrate weights and produce output spikes. The synaptic connection between neurons is realized inside each core, with a weight value associated to each connection (dependent on the programmed axon type). Each core is formed by 256 axons and 256 neurons, totaling $2^{16}$ configurable synapses per core \cite{Merolla_etal14}.}
\label{fig:tn_chip}
\end{center}
\end{figure*}

\subsection{Converting to a Spiking Neural Network}
The conversion of the original FCNN to a spiking network has to address the following issues. 
First, it needs to substitute the 32-bit precision information transmission with single bit precision. 
Second, the information is not transmitted at every time step but instead only when enough information is acquired by the processing unit (i.e. when the neuron's membrane potential crosses its threshold).

These differences in information transmission imply in a range of changes to the original architecture. 
The first issue is easily addressed by only sending "spikes", i.e. the single bit information packet. 
This discretization will lead to high error if it is not compensated for.
Two common ways of dealing with this discretization is to either use many units that send the same information to increase precision (population coding), or to use more than one time step to transmit the information (rate coding). In both cases, there needs to be a mechanism to determine when transmission of the next input pattern begins.

\begin{figure*}
\centering
		\includegraphics[height=5cm]{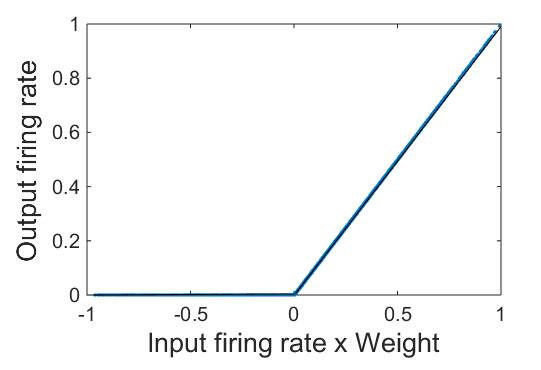}
		\includegraphics[height=5cm]{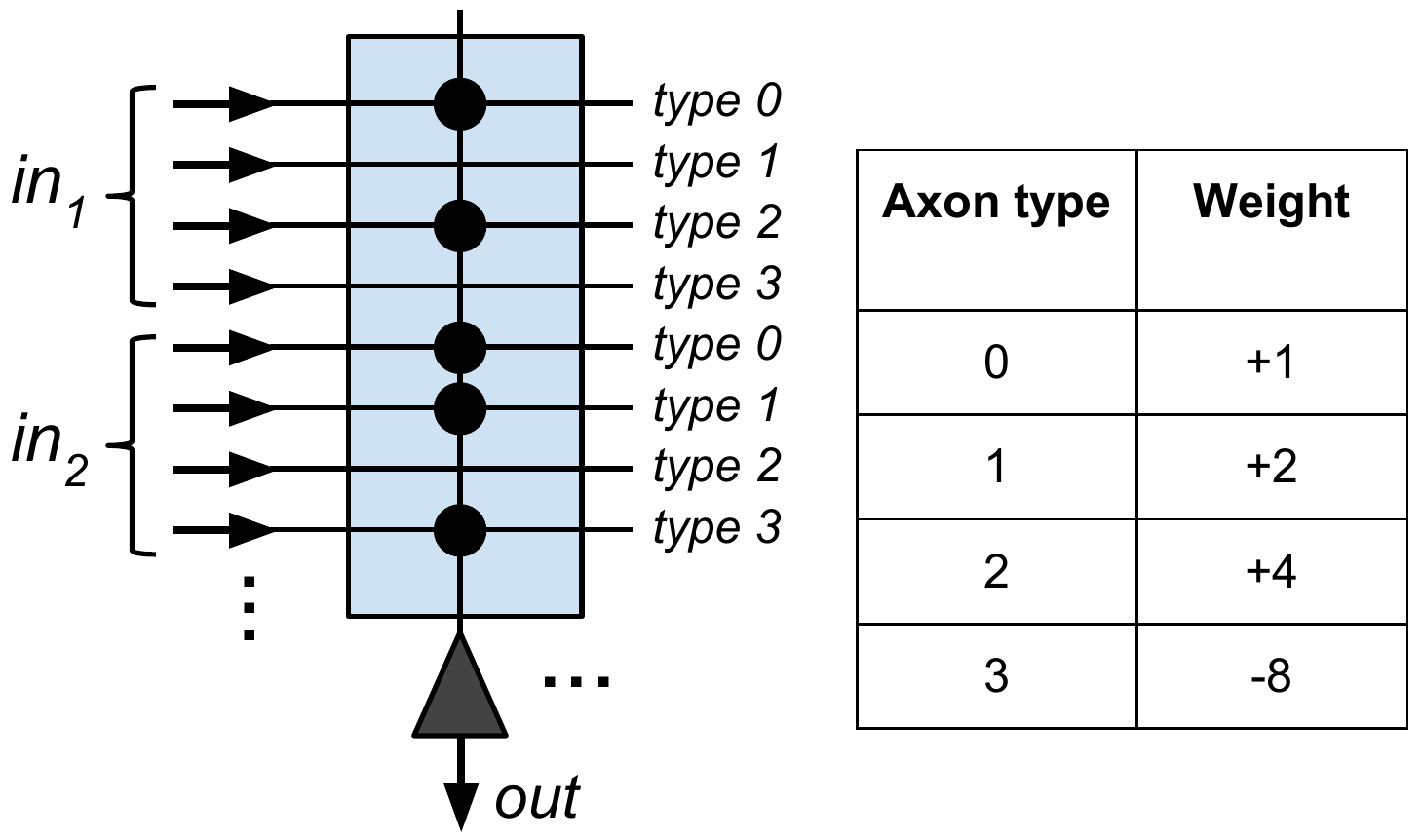}
\caption{ Rectified linear unit (ReLU) implemented using TrueNorth neurons (left). Example of a TrueNorth crossbar with multi-axon strategy per input for better weight representation per synaptic connection (right). }
\label{fig:4axon}
\end{figure*}

The second issue, in theory, can be easily addressed by setting the spiking threshold as small as the smallest incoming weight. 
However, in practice the amount of information which is integrated before a spike is transmitted should be taken into account. 
In other words, the relation between the "spiking threshold" of the units and the magnitude of the incoming weights must be considered to reduce the loss of information when many spikes arrive at the neuron during one timestep \cite{Diehl_etal15}.
Here we decided to substitute the ReLUs with integrate-and-fire neurons since both units' response is linearly proportional to the input signal and is zero if the sum of the input is negative, see Figure \ref{fig:4axon} (left).
For a better representation of the input we chose to accumulate inputs over time.
Specifically, we use spike rate code with a Poisson spike-train, where at each time step the probability of an input spike is proportional to the strength of the input.
One caveat of the system involves negative inputs, since there is no direct way of representing them using only an "on" signal.
We present two approaches to address this problem.
The first consists of doubling the number of input neurons, copying the weights from the trained network and then multiplying them by -1.
If the number of input neurons is an issue, which is commonly the case in neuromorphic systems, it would be more useful to transform the input such that it is non-negative.
For our TrueNorth implementation, we chose to follow a second strategy, where we trained and tested the network using the exponential transform of the original input, thereby assuring that all input values are non-negative.

\begin{figure*}
\centering
		\includegraphics[height=6cm]{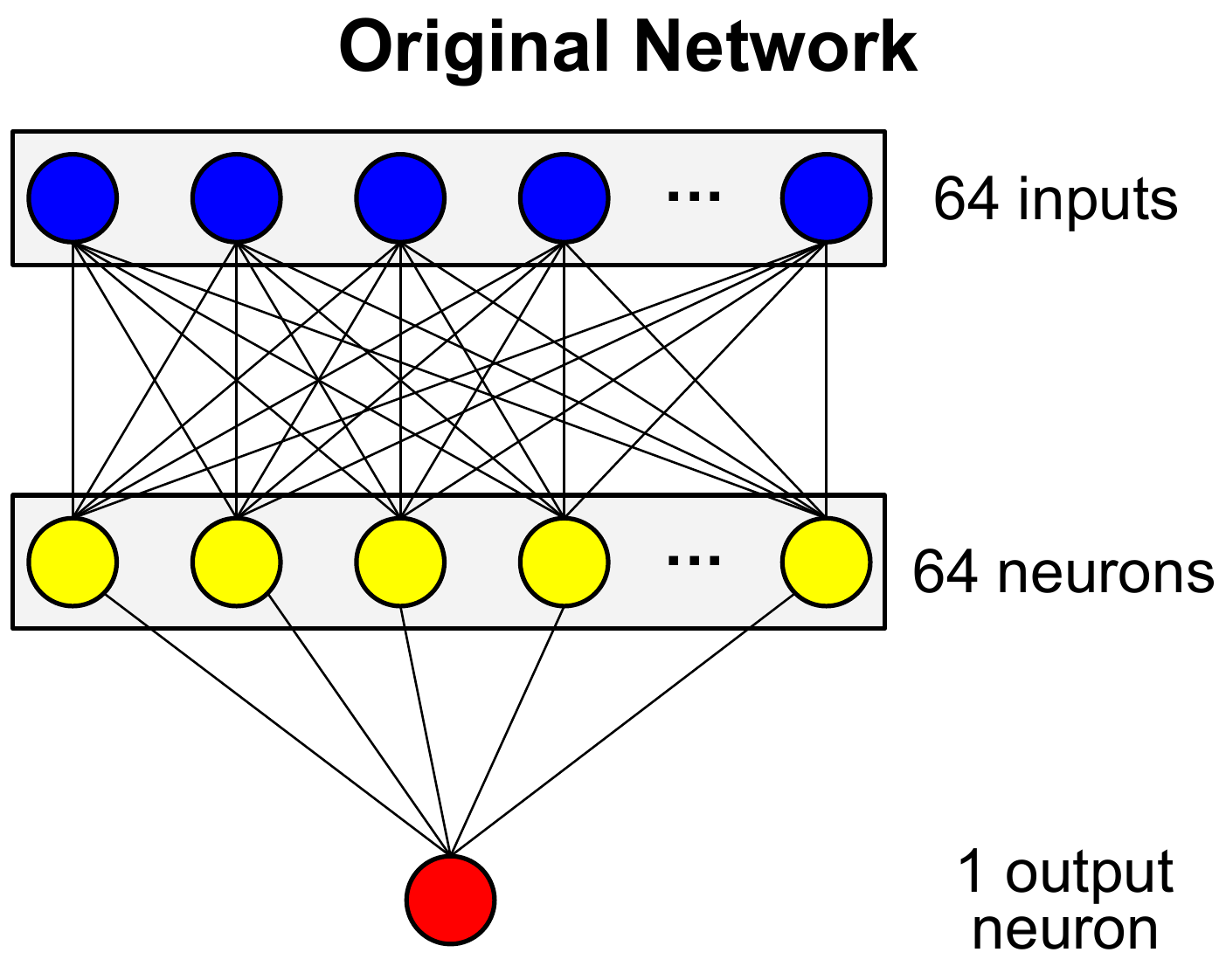}
		\includegraphics[height=6cm]{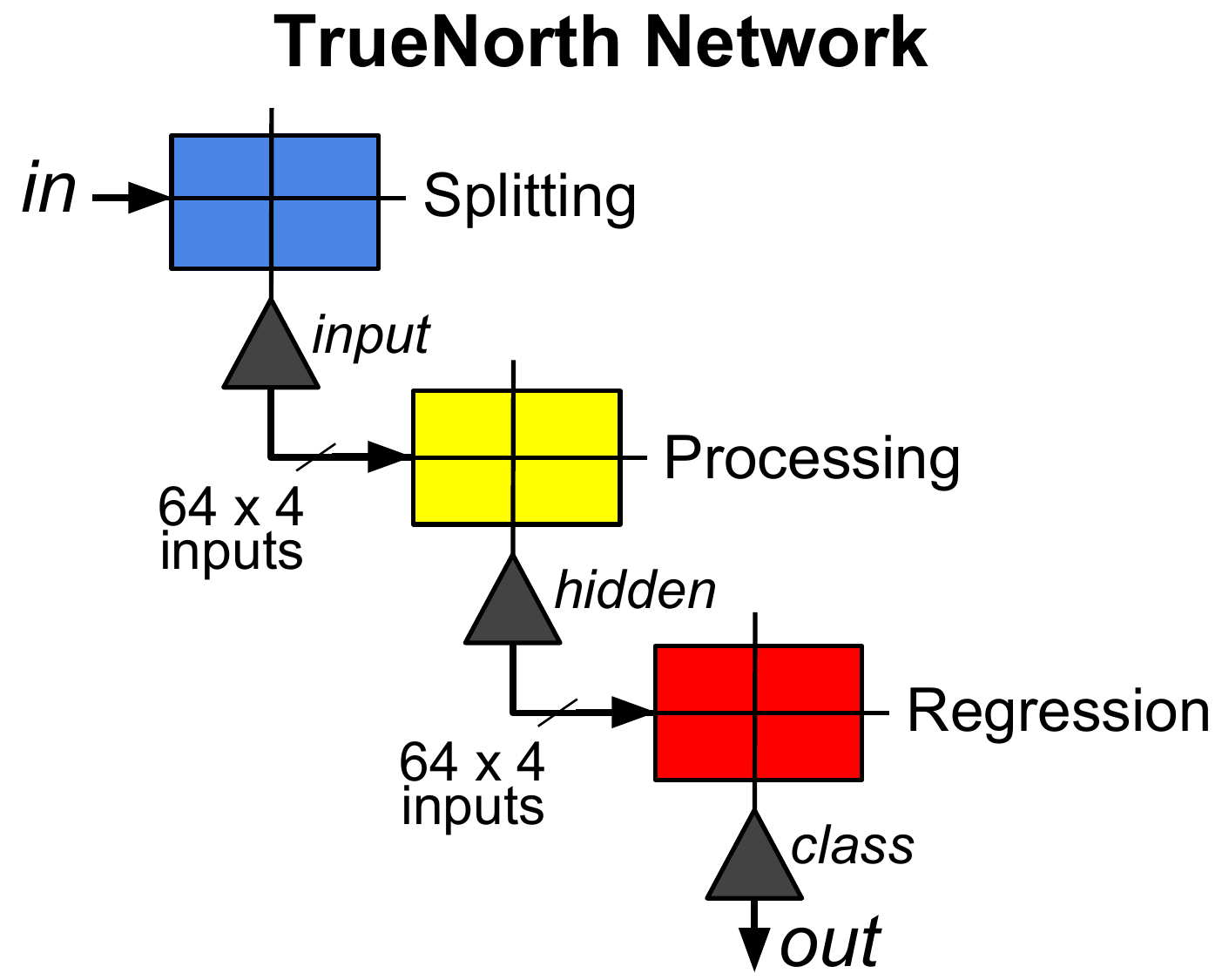}
\caption{Structure of the fully-connected neural network and the corresponding implementation on TrueNorth using 3 out of the 4096 cores available on one chip.}\label{fig:network}
\end{figure*}

\subsection{TrueNorth}\label{subsec:tn}
The IBM TrueNorth is a very low power digital neurosynaptic processor that uses brain-inspired processing and topology \cite{Merolla_etal14, Merolla_etal11}. 
The chip consists of 1 million programmable spiking neurons and 256 million configurable synapses, uniformly distributed throughout a 64 x 64 core architecture. 
Each core is composed of 256 axons (inputs) and 256 neurons (outputs), connected via a 256 x 256 crossbar of configurable synapses (see Figure \ref{fig:tn_chip}). 
The system operates in 1 ms timesteps (``ticks''), during which membrane potential processing and spike event routing occur asynchronously inside the chip. 
Spikes generated by a neuron can target any single axon on the chip, with each neuron presenting over 20 individually programmable features (e.g. threshold, leak, and reset).

The following three sequentially processed equations define the dynamics of the membrane potential $V_{j}(t)$ for neuron $j$ at time $t$ (for the full equation see \cite{cassidy2013}:

\begin{equation}
V_{j}(t) = V_{j} (t-1) +\Sigma_{i=0}^{255} \ A_{i}(t) \ w_{i,j} \ s_{j}^{G_i} \label{eq:tn1}\\
\end{equation}
\begin{equation}
V_{j}(t) = V_{j}(t) + \lambda_{j} \label{eq:tn2} \\
\end{equation}
\begin{equation}
\textrm{if} \ (V_{j}(t) \ge \alpha_{j}), \ \textrm{Spike and set } V_{j}(t) = V_{j}(t)-\alpha_{j} \ \ \ \textrm{   } \ \ \ \label{eq:tn3}
\end{equation}

The first equation represents the synaptic integration of all active axons at time $t$. 
The term $A_i(t)$ is the binary-valued input spike on the $i$-th axon at time $t$; $w_{i,j}$ is the binary-valued synaptic connection between axon $i$ and neuron $j$; and $s_{j}^{G_i}$ is the synaptic weight between axon $i$ and neuron $j$. 
Each neuron presents four individually configurable 9-bit signed integer weights, however the last term in equation \ref{eq:tn1} is dependent on the axon type, making each connection's weight not independent. 
In this manner, an axon can be configured to be one of four types, and this defines which of the four weights (in each neuron) will be integrated into each neuron case the axon is active. 

The second equation simply integrates the configurable leak value for the neuron. 
Lastly, the third equation compares the updated membrane potential (after weight integration) with the threshold. 
One of the neuron properties particular to TrueNorth is the ``linear reset mode''.
In this mode, if $V_{j}(t)$ is equal to or surpasses the threshold ($\alpha_{j}(t)$), the neuron spikes and its membrane potential is subtracted by the threshold value. 

Given the high-level view of the TrueNorth system, the following subsection will cover the implementation details of the neural network mapped onto TrueNorth, including the architectural and algorithmic adaptations demanded when using digital spiking neurons for inference \cite{amir2013}.

\begin{figure*}
\begin{center}
\includegraphics[width=17.5cm]{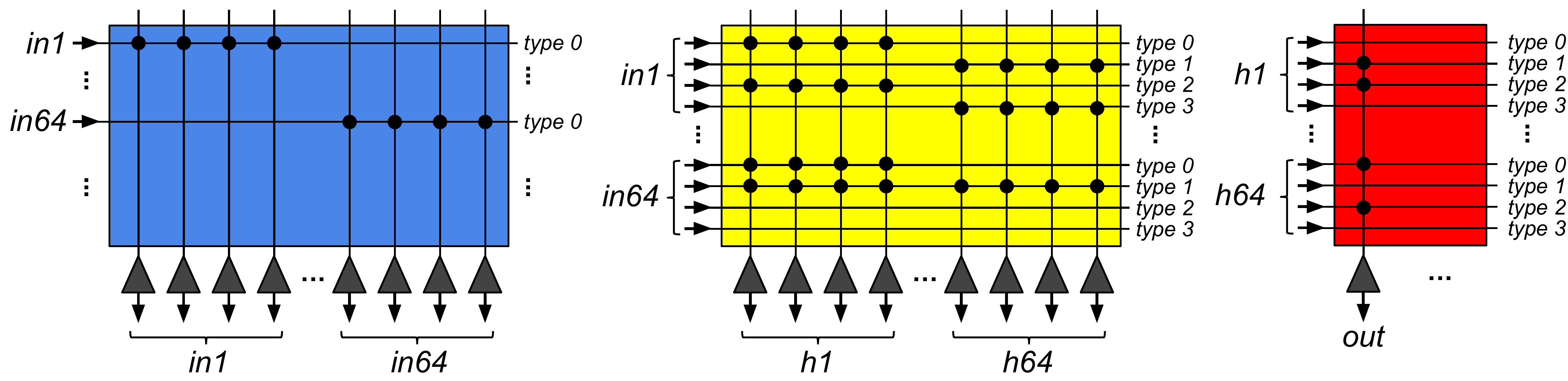}
\caption{Example configuration of the crossbars on TrueNorth cores for each network layer. The first core is used for replicating the inputs for better weight representation (left). The second core processes spikes between the input layer and the hidden layer (center). The third core processes spikes between the hidden layer and the output neuron, where the number of output spikes represents the winner class (right).}
\label{fig:tn_network}
\end{center}
\end{figure*}

\subsection{Implementation Details}\label{sec:implementation_details}
The architecture of the original artificial neural network to be mapped onto the TrueNorth system can be seen in Figure \ref{fig:network} (left). 
The entire network was mapped using 3 TrueNorth cores -- one per layer. 
A high-level architecture of the TrueNorth network is shown in Figure \ref{fig:network} (right). 
The first layer consists of 64 inputs: the 64-dimensional word2vec representation of the input word under analysis. 
The next layer is comprised of 64 hidden units, all connected to a single output regression neuron.

The TrueNorth implementation of the FCNN uses the linear reset mode to closely match the ReLUs of the original network and 4-axon inputs to obtain better weight representation.
Figure \ref{fig:4axon} (left) shows how a linear reset TrueNorth neuron configured with threshold $\alpha=100$ and $\lambda=0$ behaves similarly to a ReLU: there is a linear relationship between ``normalized product of input firing rate and connection weight'' and the ``normalized output firing rate''.

The 64 inputs in the second and third layers (cores) make use of all of the 256 axons on their respective cores, while the first core is used simply for splitting (replicating $4\times$) the inputs.
The reason for using all 256 axons on one core for 64 inputs is that the weight precision on TrueNorth is limited due to the axon type defining the weight used in each connection.
A strategy for obtaining a better weight representation in TrueNorth is to use multiple axons for each input. 
In this manner, the total effect of an active input will be a linear combination of all the axons it is connected to, resulting in multiple available weights per connection. 
Figure \ref{fig:4axon} (right) visualizes this strategy, where 4 axons are used per input, resulting in $2^4$ different possible weight values available per input-output connection.   
In the example, the weight between $in_1$ and $out$ equals +5, while the weight between $in_2$ and $out$ equals -5.
The main drawback is the need to use ``splitter'' cores for replicating a single input to multiple axons. 

Figure \ref{fig:tn_network} (left) shows how the neurons and axons are used for each layer.
The first core is used to create replicas of the 64 inputs, which is obtained by setting the neurons with threshold $\alpha=1$ and making the weight of all connections also equal to one. 
In this manner, any incoming spike event will trigger 4 replicas of this event. The second core, shown in Figure \ref{fig:tn_network} (center), is used to process the input spikes by integrating the mapped weights and generating hidden layer spike events whenever the membrane potential crosses its spiking threshold. 
The final core is depicted in Figure \ref{fig:tn_network} (right) and operates in the same manner as the second core, however now the spiking activity of the output neuron is sent off-chip. 
The regression task, therefore, produces a result based on the number of spikes which were output during the experiment time window $T$.

The data, code for the Python implementation, and code for the real-time TrueNorth demo will be made publicly available but the latter will require access to a TrueNorth chip or to the Compass simulator (for which currently an agreement with IBM is required).

\section{Results}
After training the fully-connected network on 4000 word vector and sentiment pairs, we test the sentiment prediction performance on the remaining 500 word-sentiment pairs of the dataset. 
We measure the Pearson correlation between the human sentiment annotations and our networks' predictions. 
Table \ref{tab:performance} shows the performance for the original DNN and for the converted SNN using a range of different setups.
The original DNN and the DNN trained with exponentiated inputs both achieve a fairly high correlation above 0.6.
When discretizing the weights of the networks, the correlation goes down to 0.499 and 0.408 for the original and the exponential inputs, respectively.
The SNNs derived from these four different networks show comparable performances. 
Notably, as integration time increases, the performance of the SNNs approaches the performance of their 32-bit counterpart for every single one of the four setups.

\begin{table}
\begin{center}
\begin{tabular}{ll} \hline
Configuration &	Correlation \\ \hline
Original DNN &  \textbf{0.637} \\
SNN, 0.01s integration &	0.291 \\
SNN, 0.10s integration &	0.377 \\
SNN, 1.00s integration &	\textbf{0.631} \\ \hline
Original DNN - exp&  \textbf{0.661} \\
SNN, 0.01s integration - exp &	0.279 \\
SNN, 0.10s integration - exp &	0.544 \\
SNN, 1.00s integration - exp &	\textbf{0.651} \\ \hline
Original DNN - 4 bit&  \textbf{0.499} \\
SNN, 0.01s integration - 4 bit &	0.183 \\
SNN, 0.10s integration - 4 bit &	0.293 \\
SNN, 1.00s integration - 4 bit &	\textbf{0.486} \\ \hline
Original DNN - exp, 4 bit&  \textbf{0.408} \\
SNN, 0.01s integration - exp, 4 bit &	0.310 \\
SNN, 0.10s integration - exp, 4 bit &	0.380 \\
SNN, 1.00s integration - exp, 4 bit &	\textbf{0.392} \\ \hline \vspace{9pt}
\end{tabular}\caption{Results for sentiment test set. The table shows the correlation between the predicted sentiment and the correct sentiment labels for the original model and the SNN. With increasing integration time the performance of the SNN approaches the performance of the original rate-based network.} \label{tab:performance}
\end{center}
\end{table}

\begin{figure}
\begin{center}
\includegraphics[width=8.5cm]{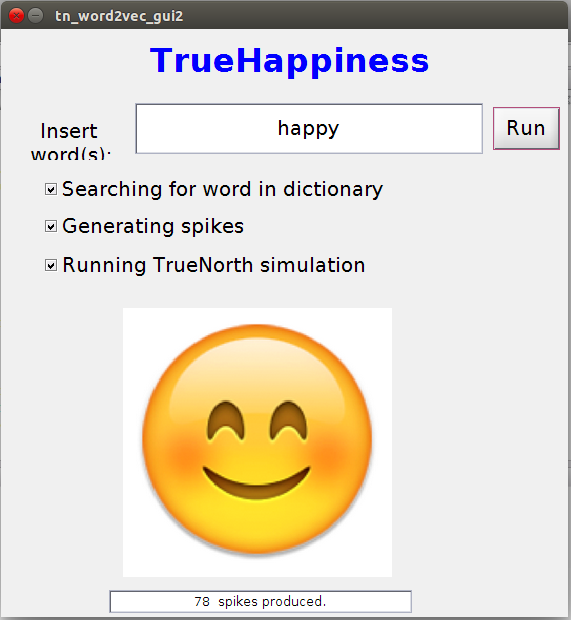}
\end{center}
\caption{The interface for the real-time interactive demo. The user can type words contained in the $>$300,000 word vocabulary and the system will predict its sentiment using TrueNorth.}
\label{fig:demo-interface}
\end{figure}

TrueHappiness, the sentiment prediction system adapted for TrueNorth, uses an interactive GUI where the user can type any word in the $>$300,000 word vocabulary. 
The input word is then converted to a constant length vector using word2vec, which is then converted to Poisson spike trains and transmitted to the TrueNorth chip.
On TrueNorth, the SNN processes the input and sends the output spikes back to the computer where the sentiment is categorized into classes based on the number of spikes, and the class and the number of output spikes are displayed in the GUI (see Figure \ref{fig:demo-interface}).

\section{Discussion}
We showed a first example of a low power NLP system by leveraging established machine learning techniques and mapping the resulting neural networks to the TrueNorth chip.
This framework for converting fully-connected networks trained with backpropagation to a TrueNorth compatible network is more generally applicable. 
In particular, the framework offers high flexibility because of its modularity.
A neural network designer can optimize the performance of the network by improving optimization methods, loss functions or other recently developed methods and, as long as the weights and connectivity are in the required format, the rest of the processing pipeline does not change. 
Similarly, the rest of the pipeline is oblivious to which library is used for training the NN, which means the NN designer has the freedom to choose libraries such as Pylearn2/Theano \cite{Bergstra_etal10}, Torch or Caffe \cite{jia2014}.
Moreover, none of the steps in the framework (besides the word2vec preprocessing) depend on the task to be solved. 
For example, a similar approach has been used for question classification \cite{Diehl_etal16b}.
The inputs might just as well be pixel intensities or an audio signal, without needing to change the framework.

One important issue that is not tackled here is the connectivity of the NN and it is assumed that the NN designer takes care of the connectivity constraints on TrueNorth, i.e. every neuron can only connect to one other axon (i.e. 256 neurons) and every neuron can only receive up to 256 inputs.
While this might sound like a harsh restriction, many high performance NNs, like convolutional neural networks \cite{LeCun_etal98}, use local connectivity by design and are therefore easier to adapt to TrueNorth.
Another restriction is that learning is done offline, i.e. the system can not adapt its parameters to new incoming data.
While this is probably acceptable for many practically relevant systems, many other systems might need to adjust to a specific user or a new environment.
In such cases it would be necessary to use neuromorphic hardware that brings efficient online learning \cite{Khan_etal08, diehl2014, galluppi2014} in combination with online learning algorithms \cite{beyeler2013, Masquelier_etal09, Kappel_etal14, Neftci_etal13a, Diehl_Cook15}. 
However, so far it remains a challenge to achieve performances comparable to conversion methods using such online learning approaches.

The mapping from the FCNN to the spiking network does lead to a performance gap between the original network and its spiking counterpart.
However, this gap is very small for longer presentations of the inputs and makes it possible for the system designer to trade-off accuracy for resources depending on the application.
For example in an application where there is very high variance in the inputs, it is likely not useful to push the limits of accuracy for a given input but instead decrease latency, thus increasing throughput. 
Another way to minimize the performance gap, would be to duplicate the inputs and thereby use multiple Poisson spike-trains as inputs per actual input.
Since those two methods are complementary, they can also be combined.  

A more important source of performance loss is the weight discretization to 4 bits which is used here for the mapping to TrueNorth.
Using the naive approach of discretizing or rounding the weights to the nearest of the 16 values results in a reduction in Pearson correlation of 0.138.
While this is quite a significant reduction, there are many ways to improve upon this method.
One possibility is to use a layer-wise training method where after the first rounding step the resulting network is trained again (using standard backpropagation) but with fixed weights for the first layer.
After this re-training, the weights are rounded again and the re-training could then be repeated as often as there are remaining layers.
In this way, the network will learn to use the discretized weights in the earlier layers in a better way and potentially compensate the discretization loss.
Other ways would be to use stochastic rounding \cite{gupta2015deep} or more sophisticated training methods such as the dual-copy rounding \cite{stromatias2015}, where the weights are discretized in the forward pass of the backpropagation algorithm but full resolution weights are used for backpropagating the error signal. 
This results in weights that show only minimal loss due to discretization.
If resources are not a major issues at run time, it would also be possible to use a probabilistic rounding of the weights and use multiple instances of the network.
For example if a weight has a value of 0.7 and we want to discretize is to 0 or 1, we could flip ten biased coins, using every coin for one of ten different networks. 
A possible outcome of those coin flips would be that seven of the ten networks are using a weight of 1 and the remaining three networks use a weight of 0.
By averaging over the results of these networks, the result will most likely improve, similar to other committee methods in machine learning \cite{cirecsan2011}, but only requiring to train the original network once.
"Constrain-then-train" approaches \cite{Esser_etal2015} successfully employ strategies such as these for maximizing runtime system accuracy.

In order to deal with negative inputs, we introduced two methods for representing these data using spike trains.
The first one consists of duplicating every input neuron which potentially needs to represent a negative value.
This leads to a very small performance gap between the original and the spiking network, i.e. the correlation only decreased by 0.006.
The other presented approach is to exponentiate the inputs to ensure that they are non-negative.
Interestingly, for the sentiment analysis task this even improved the performance of the original network as well as the performance of the spiking network, even though the performance gap slightly increased to 0.01.
After discretizing the weights to 4 bits to be able to map them on TrueNorth, the performance gap increases slightly for both methods: to 0.013 for the input duplication and to 0.016 for the exponentiation. 
However, in the case of the exponentiation, the drop in correlation due to weight discretization is with 0.253 much bigger than the performance gap for spiking networks, rendering the performance gap due to a spiking representation relatively insignificant.
A likely explanation for this increased loss is that the exponentiation reduces the absolute difference between some words and increases the difference for other words.
This would require more finely tuned weights to correctly process the small differences, but this  exact fine tuning will not be represented in the 4-bit weights.
It remains an open question whether the above mentioned methods for reducing the discretization loss might help to avoid the negative effect of the exponentiation.

Despite the mentioned challenges, this work represents an important first step towards practically relevant and extremely low power NLP systems and other neuromorphic pattern recognition systems.
The main advantage of our method lies in its modularity and flexibility since progress in deep learning research will directly impact the performance of neuromorphic recognition systems designed with our framework.

\section*{Acknowledgments}
We thank the organizers and the participants of Telluride Neuromorphic Cognition Engineering Workshop 2015 for the unique environment that enabled the presented work. Especially the natural language processing group, Rodrigo Alvarez, and John Arthur for many fruitful discussions.

\textit{Funding:} 
PUD: SNF Grant 200021-143337 "Adaptive Relational Networks."
BUP: The Office of Naval Research (ONR MURI 14-13-1-0205) and CNPQ Brazil (CsF 201174/2012-0)
EN: the Office of Naval Research (ONR MURI 14-13-1-0205)




%

\bibliographystyle{IEEEtran}
\bibliography{biblio}

\end{document}